\documentclass[runningheads]{llncs}
\usepackage{graphicx}
\usepackage{amssymb}

\usepackage{amsmath}
\DeclareMathOperator*{\argmax}{arg\,max}

\usepackage{mathtools}
\DeclarePairedDelimiter{\ceil}{\lceil}{\rceil}
\DeclarePairedDelimiter{\floor}{\lfloor}{\rfloor}

\newtheorem{observation}{Observation}

\usepackage[parfill]{parskip}

\usepackage{wrapfig}

\usepackage{hyperref}

\usepackage{enumerate}

\usepackage{color}

\newcommand{\rsec}[1]{Section~\ref{sec:#1}}
\newcommand{\rthm}[1]{Theorem~\ref{thm:#1}}
\newcommand{\rlem}[1]{Lemma~\ref{lem:#1}}

\newcommand{\rfig}[1]{Figure~\ref{fig:#1}}
\newcommand{\ropt}[1]{Option~\ref{opt:#1}}

\newcommand{\Oh}{\mathcal{O}}
\newcommand{\opt}[2]{#1[#2]}
\newcommand{\h}{h}
\newcommand{\conf}[2]{C_{#1 \hookrightarrow #2}}

\usepackage{color}

\newcommand{\GR}{\textsc{Gread}}
\newcommand{\OFF}{\emph{OFF}}
\newcommand{\ON}{\emph{ON}}
\newcommand{\dist}{\emph{d}}

\newcommand{\mtf}{\textsc{mtf}}
\newcommand{\pwl}{\textsc{Distributed List Update}}
\newcommand{\dga}{\textsc{Distributed Grid Update}}

\newcommand{\ALG}{\textsc{A}}
\newcommand{\Cost}{\text{cost}}
\newcommand{\lus}{\tau}

\usepackage{xcolor}
\colorlet{red}{black}

\begin{document}
\title{Self-Adjusting Linear Networks}
%
%
\author{Chen Avin\inst{1} \quad
Ingo van Duijn\inst{2} \quad
Stefan Schmid\inst{3}}
\authorrunning{Avin et al.}
%
\institute{Ben Gurion University of the Negev \\
\and Aalborg University \\
\and University of Vienna}
%
\maketitle              
\begin{abstract}
Emerging networked systems become increasingly flexible
and ``reconfigurable''. This introduces an opportunity
to adjust networked systems in a demand-aware manner,
leveraging spatial and temporal locality in the workload for \emph{online} optimizations. 
However, it also introduces a tradeoff: while more frequent
adjustments can improve performance, they also entail 
higher reconfiguration costs. 

This paper initiates the formal study of
\emph{linear} networks which self-adjust to the demand
in an online manner, striking a balance between the benefits
and costs of reconfigurations. We show that 
the underlying algorithmic problem can be seen as a 
distributed generalization of the classic dynamic list update problem
known from self-adjusting datastructures: in a network, 
requests can occur between \emph{node pairs}.
This distributed version turns out to be significantly harder than the classical problem in generalizes.
Our main results are a $\Omega(\log{n})$ lower bound on the competitive ratio,
and a (distributed) online algorithm that is $\Oh(\log{n})$-competitive if the communication requests are issued according to a linear order.

\end{abstract}
\section{Introduction}\label{sec:intro}

Communication networks are becoming increasingly flexible, along three main dimensions:
routing (enabler: software-defined networking), embedding (enabler: virtualization), 
and topology (enabler: reconfigurable optical technologies, for example~\cite{projector}). 
In particular, the possibility to quickly reconfigure communication networks, e.g., by migrating
(virtualized) communication endpoints~\cite{disc16} or by reconfiguring the (optical) topology~\cite{ccr18san}, allows these networks
to become \emph{demand-aware}: i.e., to adapt to the traffic pattern they serve, in an online
and \emph{self-adjusting} manner. 
For example, in a self-adjusting network, frequently communicating node
pairs can be moved \emph{topologically closer}, saving communication costs 
(e.g., bandwidth, energy) and
improving performance (e.g., latency, throughput).

However, today, we still do not have a good understanding yet of the
algorithmic problems underlying self-adjusting networks. 
The design of such algorithms faces several challenges. 
As the demand is often not known ahead of time, \emph{online}
algorithms are required to react to changes in the workload in a clever way;
ideally, such online algorithms are ``competitive'' even when compared
to an optimal offline algorithm which knows the demand ahead of time.
Furthermore, online algorithms need to strike a balance between the benefits
of adjustments (i.e., improved performance and/or reduced costs)
and their costs (i.e., frequent adjustments can temporarily harm 
consistency and/or performance, or come at energy costs).
 
The vision of self-adjusting networks is reminiscent of 
self-adjusting datastructures such as 
\emph{self-adjusting lists} and \emph{splay trees}, which 
optimize
themselves toward the workload.
In particular, the \emph{dynamic list update problem},
introduced already in the 1980s by Sleator and Tarjan
in their seminal work~\cite{sleator1985amortized},
asks for an online algorithm to reconfigure 
an unordered linked list datastructure, 
such that a sequence of lookup requests is served optimally
and at minimal reconfiguration costs (i.e., pointer rotations). 
It is well-known that a simple \emph{move-to-front} strategy,
which immediately promotes each accessed element to the front
of the list, is \emph{dynamically optimal}, that is, has a constant competitive ratio.

This paper initiates the study of a most basic self-adjusting
linear \emph{network}, which can be seen as a \emph{distributed} variant
of the dynamic list update problem, generalizing the datastructure
problem to networks: while datastructures serve requests 
originating from the front of the list (the ``root'')
to access data items, networks serve \emph{communication} requests between
\emph{pairs of nodes}. 
The objective is to move nodes which currently
communicate frequently, closer to each other, 
while accounting for reconfiguration costs.

\subsection{Formal Model}

\color{red}

We initiate the study of pairwise communication problems in a dynamic network reconfiguration model,
using the following notation:
\begin{itemize}
\item Let $d_G(u,v)$ denote the \emph{(hop) distance} between $u$ and $v$ in a graph $G$.
\item A \emph{communication request} is a pair of communicating nodes from a set $V$.
\item A \emph{configuration} of $V$ in a graph $N$ (the host network) is an injection of $V$ into the vertices of $N$;
$\conf{V}{N}$ denotes the set of all such configurations.
\item A configuration $\h \in \conf{V}{N}$ is said to \emph{serve} a communication request $(u,v) \in V \times V$ at cost $d_N(\h(u), \h(v))$.
\item A finite \emph{communication sequence} $\sigma = (\sigma_0, \sigma_1,\ldots, \sigma_m)$ is served by a sequence of configurations $\h_0, \h_1,\ldots,\h_m \in \conf{V}{N}$.
\item The cost of serving $\sigma$ is the sum of serving each $\sigma_i$ in $\h_i$ plus the reconfiguration cost between subsequent configurations $h_i, h_{i+1}$.
\item The reconfiguration cost between $h_i, h_{i+1}$ is the number of \emph{migrations} necessary to change from $h_i$ to $h_{i+1}$;
a migration swaps the images of two nodes $u$ and $v$ under $\h$.
\item $E_i = \{\sigma_1,\ldots,\sigma_i\}$ denotes the first $i$ requests of $\sigma$ interpreted as a set of edges on $V$,
and $R(\sigma) = (V, E_m)$ denotes the \emph{request graph} of $\sigma$.
\end{itemize}

In particular, we study the problem of designing a self-adjusting \emph{linear network}:
a network whose topologoy forms a $d$-dimenstional grid.
We are particularly interested in the 1-dimensional grid in this paper, the line:
\begin{definition}[\pwl]
Let $V$, $\h$, and $\sigma$ be as before, with
\[N = (\{1,\ldots,n\}, \{(1,2), (2,3), \ldots, (n-1, n)\}\]
representing a list network.
The cost of serving a $\sigma_i=(u,v) \in \sigma$ is given by $|h(u) - h(v)|$, i.e.\ the distance between $u$ and $v$ on $N$.
Migrations can only occur between nodes configured on adjacent vertices in $N$.
\end{definition}
\color{black}

Recall that the cost incurred by an algorithm $\ALG$
on $\sigma$ is the sum of communication and reconfiguration costs.
In the realm of online algorithms and competitive
analysis, we compare an online algorithm $\ON$
to an offline algorithm $\OFF$ which has complete knowledge
of $\sigma$ ahead of time.
We want to devise online algorithms $\ON$ which minimize
the competitive ratio $\rho$:
$$
\rho = \max_{\sigma} \frac{\Cost(\ON(\sigma))}{\Cost(\OFF(\sigma))}
$$
\textcolor{red}{
As a first step, we in this paper consider the \pwl{} problem for the case where 
the request graph $R(\sigma)$ has constant \emph{graph bandwidth}:
i.e.\ graphs for which there is a configuration in a line network such that any request can be served at constant cost.
We refer to such a request graph as \emph{linear demand}.
}
\subsection{Contributions}

This paper initiates the study of a most basic self-adjusting
network, a line,  which optimizes itself toward the dynamically changing
linear demand, while amortizing reconfiguration cost. 
The underlying algorithmic problem is natural and motivated by
emerging reconfigurable communication networks (e.g., based
on virtual machine migration or novel optical technologies~\cite{disc17,projector}).
The problem can also be seen as a 
distributed version of the fundamental
dynamic list update problem. 
Our first result is a negative one:
we show that unlike the classic dynamic list update problem,
which admits for constant-competitive online algorithms,
there is an $\Omega(\log{n})$ lower bound on the competitive ratio of any deterministic 
online algorithm for the distributed problem variant. 
\textcolor{red}{Our second main contribution is 
a (distributed) online algorithm which is 
$\Oh(\log{n})$-competitive for long enough sequences. }

\subsection{Organization}

The remainder of this paper is organized as follows.
In Section~\ref{sec:generalization}, we put the problem and its challenges 
into perspective with respect to the list
update problem.
We then first derive the lower bound in Section~\ref{sec:lowerbound}
and present our algorithm and upper bound in Section~\ref{sec:upperbound}.
After discussing related work in Section~\ref{sec:relwork},
we conclude in Section~\ref{sec:conclusion}.

\section{From List Update to Distributed List Update}\label{sec:generalization}

To provide an intuition of the challenges involved in 
designing online algorithms for distributed list update problems
and to put the problem into perspective, we first revisit
the classic list update problem and then discuss why similar
techniques fail if applied to communicating node \emph{pairs}, i.e.,
where requests not only come from the front of the list.

The \emph{(dynamic) list update problem}~\cite{sleator1985amortized} 
introduced by Sleater and Tarjan over 30 years ago is
one of the most fundamental and oldest online problems:
Given a set 
of $n$ elements stored
in a linked list, how to update the list over time
such that it optimally serves a request sequence $\lus =(\lus_1,\lus_2,\ldots)$ where for each $i$, $\lus_i \in V$ is an arbitrary element
stored in the list? 
The cost incurred by an algorithm is the sum of the access costs
(i.e.\ scanning from the \emph{front} of the list to the accessed element) and
the number of \emph{swaps} (switching two neighboring elements in the list).
As accesses to the list elements start at the front of the list,
it makes sense to amortize high access costs by moving frequently accessed elements
closer to the front of the list.
In fact, the well-known \emph{Move-To-Front} (MTF) algorithm even moves an accessed element to the front
\emph{immediately}, and is known to be \emph{constant competitive}:
its cost is at most a factor 2 (or some other constant, depending on the cost model) worse than that of
an optimal offline algorithm which knows the entire sequence $\lus{}$
ahead of time \cite{sleator1985amortized}.
Throughout the literature, slightly different cost models have been used for the 
list update problem, though they only differ by a constant factor.
Generally, a \emph{cursor} is located at the head of the list at each request.
Then, the algorithm can perform two operations, each operation incurring unit cost.
i) \emph{Move} the cursor to the left, or to the right, one position; 
the element in the new position is referred to as \emph{touched}.
ii) \emph{Swap} the element at the cursor with the element one 
position to the left or right; the cursor also moves.

In the \pwl{} problem, upon a request $\sigma_i = (s_i, t_i)$, the cursor is placed at $s_i$ 
instead of the head of the list, and $t_i$ needs to be looked up.
To demonstrate the significance of this difference, we first present a paraphrased version of the proof by Tarjan and Sleator showing the dynamic optimality of \mtf{}.
After that, we showcase a simple access sequence differentiating the two problems.

\subsection{An Expositional Proof for the Optimality of MTF}

\begin{wrapfigure}{r}{4cm}
\includegraphics[width=3.3cm]{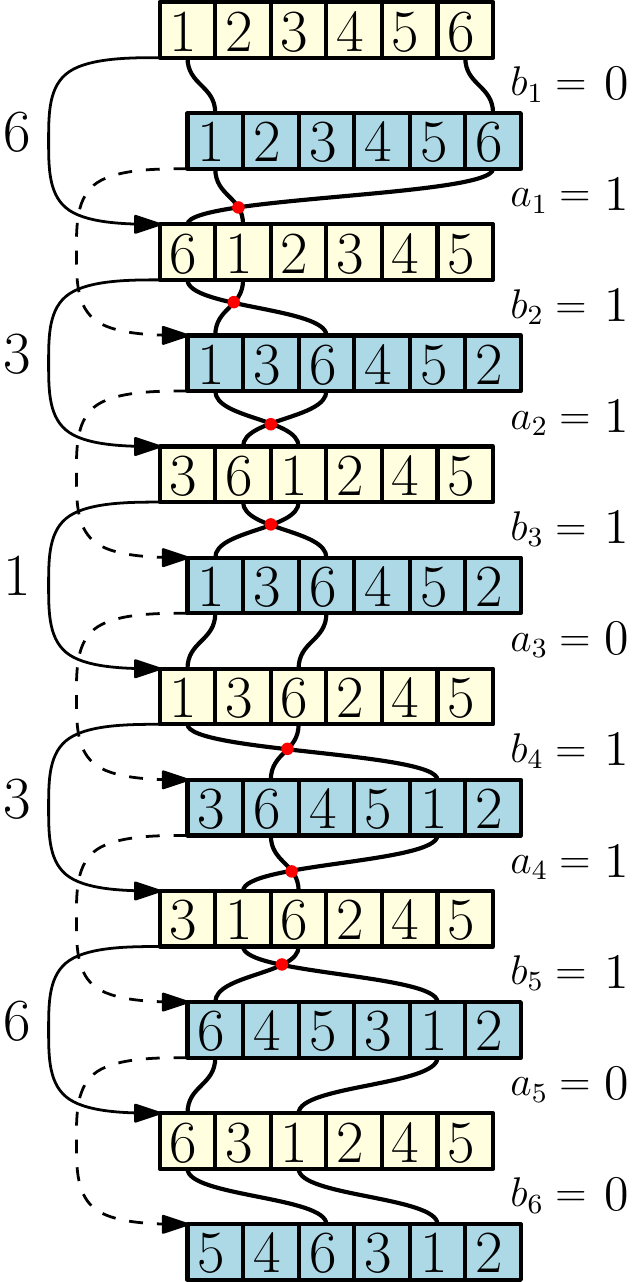}
\caption{\mtf{} (yellow) and $A$ (blue) on $\lus = 6, 3, 1, 3, 6$}
\label{fig:juxtaposition}
\end{wrapfigure} 

While the potential argument used to show dynamic optimality of the move-to-front 
strategy for the list access problem yields a very elegant and succinct proof \cite{sleator1985amortized}, it lacks intuition which makes 
it difficult to generalise the argument.
The key idea in the potential argument is to compare the execution of \mtf{} to the execution of an arbitrary algorithm $A$.
The algorithm is fixed for the analysis, but any valid algorithm can be used, e.g.\ the optimal offline algorithm.
The state (represented by a list) of \mtf{} and $A$ are juxtaposed at every access, comparing how the order of elements in both lists differ.
In fact, it is sufficient to only considers the relative order of two fixed elements $u$ and $v$ as follows.
Consider the order of $u$ and $v$ in the state of $A$ before it performs the $i$th access.
If this order is the same as in \mtf{} \emph{before} it performs the $i$th acces, let $b_i = 0$ and otherwise $b_i = 1$.
Similarly, if the relative order is the same in \mtf{} \emph{after} its $ith$ access, let $a_i = 0$ and otherwise $a_i = 1$.
This describes an inversion sequence $b_1a_1b_2a_2\dots b_ma_m$.
\rfig{juxtaposition} illustrates this for \mtf{} and an arbitrarily chosen algorithm $A$ on a sequence $\lus = 6, 3, 1, 3, 6$, with the inversions of $1$ and $6$ described by the sequence $01111011100$.

Suppose that $\lus_i \in \{u,v\}$ and that \mtf{} touches $u$ and $v$ while accessing $\lus_i$.
The proof by Tarjan and Sleator boils down to three observations.

\begin{observation}
\label{obs:invert}
MTF inverts $u$ and $v$ relative to $A$ by accessing $\lus_i$, i.e.\ $b_i \not = a_i$.
\end{observation}

\begin{observation}
\label{obs:ok}
If $b_i = 0$, \mtf{} and $A$ agree on the order of $u$ and $v$ before $\lus_i$.
Since \mtf{} touches both, $A$ also touches both in order to access $\lus_i$.
\end{observation}

\begin{observation}
\label{obs:charge}
For $b_i = 1$, let $j < i$ be the largest index such that $b_j = 0$ or $a_j = 0$ (note that $j$ exists because $b_1 = 0$).
When $a_j = 0$, and thus $b_{j+1} = 1$, $A$ inverts $u$ and $v$ and therefore must have touched both.
When $b_j = 0$, and thus $a_j = 1$, \mtf{} inverts $u$ and $v$ and one of them is $\lus_j$.
By Observation \ref{obs:ok}, if $b_j = 0$ and \mtf{} touches $u$ and $v$ to access $\lus_j$, then A does as well.
\end{observation}

The last observation is essentially the amortised argument rephrased as a charging argument.
We can now easily prove the dynamic optimality of \mtf{}.

\begin{theorem}[Tarjan \& Sleator]
\mtf{} is $4$-optimal.
\end{theorem}

\begin{proof}
We prove that for all $\lus_i = v$ where \mtf{} touches $u$, there is a move by A touching $u$.
\mtf{} first moves the cursor to $\lus_i$, and then swaps $\lus_i$ to the front.
Along the way it touches $u$ twice, once with a move and once with a swap, incurring a cost of 2.

For $b_i = 0$ (resp. $b_i=1$), we use Observation \ref{obs:ok} (resp. \ref{obs:charge}) to charge the cost to A touching $u$ while accessing $\lus_i$ (resp. $\lus_j$).
By Observation \ref{obs:invert}, $b_i \not = a_i$, and thus for any $\lus_k\in \{u,v\}$ with $i < k$, the largest index $j' < k$ with $b_{j'} = 0$ or $a_{j'} = 0$ must be at least $i$, and therefore $j < i \leq j'$.
This guarantees that \mtf{} charges at most a cost of 4 to one move of A.
Since all the cost incurred by \mtf{} is charged to some move of A, the claim follows.
\qed
\end{proof}

In the original work by Tarjan and Sleator, MTF is shown to be $2$-optimal.
This is because their cost model allows accessed elements to be moved to the front `for free'.
If we allow this as well, the cursor touches $u$ only once to access $v$, resulting in a factor $2$.

\subsection{The Challenge of \pwl{}}

\begin{wrapfigure}{R}{5cm}
\center
\includegraphics[width=4.8cm]{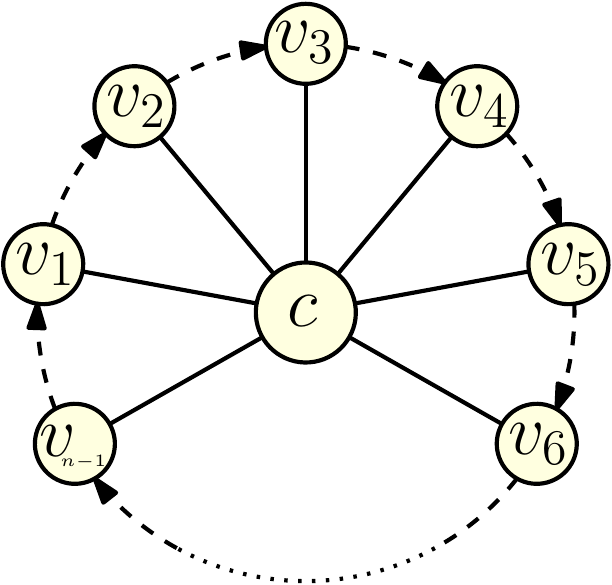}
\caption{A star graph used to construct a cyclic sequence of requests $\sigma_c = (c, v_1), (c,v_2),\dots,(c,v_{n-1}), (c,v_1),\dots$}
\label{fig:star}
\end{wrapfigure}
Generalizing dynamic list update to \pwl{} introduces a number
of challenges which render the problem more difficult.
First, the natural inversion argument no longer works: 
a reference point such as the front of the list is missing in the distributed 
setting. This makes it harder to relate algorithms to each other and 
hence also to define a potential.
Second, for general request graphs $R(\sigma)$, an online algorithm needs to be able
to essentially ``recognize'' certain patterns over time.


Regarding the latter, consider the set of nodes $V = \{v_1,...,v_{n}\}$ and let $\lus_c$ be a cyclic sequence:
for all $\lus_i, \lus_{i+1} \in \lus_c$ with $\lus_i = v_j$ and $\lus_{i+1} = v_k$ it holds that $j+1 = k (\mod n - 1)$.
From this we construct a similar sequence $\sigma_c$ for \pwl{} on the set of nodes $V \cup \{c\}$, with $\sigma_i = (c, \lus_i)$.
This yields a star graph $R(\sigma_c)$ as denoted in \rfig{star}.
An offline algorithm can clearly serve the cyclic order in optimal $\Oh(1)$ per request by moving the element $c$ one position further after every request.
However, in the list update model, any sequence cycling through all elements is a worst-case sequence.
This demonstrates that a ``dynamic cursor'' can mean a factor $n$ difference in cost.
What the sequence $\sigma_c$ also demonstrates, is that aggregating elements around a highly communicative node is suboptimal;
in the particular case of $\sigma_c$, it is this central node that needs to be moved.

\textcolor{red}{
Another pattern is a request sequence $\sigma$ that forms a connected path in the request graph $R(\sigma)$.
When restricted to only these pattersn, \pwl{} corresponds to the \emph{Itinerant List Update Problem} (ILU) studied in \cite{olver2017itinerant}.
In this work it is shown that deriving non-trivial upper bounds
on the competitive ratio already seems notoriously hard (even offline approximation factors
are relatively high).
Note that the star example can be expressed as a path, i.e.\ $\sigma'_c = (c,v_1), (v_1,c), (c,v_2),(v_2,c),(c,v_3),\ldots$,
demonstrating the significance of understanding simple request patterns for \pwl{}.
This is part of the reason why in this paper we focus on request graphs
with a linear demand.
}
\section{A Lower Bound}\label{sec:lowerbound}

This section derives a lower bound on the competitive ratio 
of any algorithm for \pwl{}.

\begin{theorem}
\label{thm:lower}
The competitive ratio $\rho = \max_{\sigma} \frac{\Cost(\ON(\sigma))}{\Cost(\OFF(\sigma))}$ for \pwl{}, with $ |\sigma|=\Omega(n^2)$, is at least $\Omega(\log n)$.
This bounds holds for arbitrarily long sequences, but if $|\sigma| = \Oh(n^2)$, it even holds if the request graph is a line.
\end{theorem}

To prove this, we consider an arbitrary online algorithm $\ON$ for \pwl{}.
The main idea is to have an adaptive online adversary construct a sequence $\sigma_\ON$ that depends on the algorithm \ON.
The adversary constructs $\sigma_\ON$ so that the resulting request graph $R(\sigma_\ON)$ is a line graph.
Because an offline algorithm knows $R(\sigma_\ON)$ in advance, it can immediately configure it and serve all requests at optimal cost of $1$.
We show that the online algorithm is forced to essentially reconfigure its layout $\log n$ times, resulting in the desired ratio.
To facilitate our analysis, we use the same notion of the \emph{distortion} of an embedding as is used in the Minimum Linear Arrangement (MLA)~\cite{diaz2002survey} problem.

\begin{definition}
Given a communication graph $G = (V,E)$ with $E \subseteq V \times V$, let $E^+ = \{(u,v) \mid d_G(u,v) < \infty\}$ denote the transitive closure of $E$.\\
For $h \in \conf{V}{N}$, let $\dist_{\h}(E)$ denote the \emph{distortion} of $E$, which is defined as:
\[\dist_{\h}(E) = \sum_{(u,v) \in E^+} \dist_{\h}(u,v)\]
\end{definition}

\color{red}
The value $\dist_{\h}(E_i)$ reflects how badly the edges in $E_i$ are configured on $N$ by $\h$.
To build $\sigma_\ON$, the adversary gradually commits to the edges of $R(\sigma_\ON)$.
Having already requested $\sigma_1,\ldots,\sigma_i$, then depending on the distortion the adversary: 
\begin{enumerate}[\textbf{Option} 1:]
    \item\label{opt:1} picks $\sigma_{i+1} = \argmax_{(u,v) \in E_i} d_\h(u, v)$.
    \item\label{opt:2} reveals a new batch of edges $M \subset V \times V$. 
\end{enumerate}

From these two options, the adversary's strategy becomes clear; \ropt{1} forces the highest possible cost to \ON{} based on $E_i$ and $\h$, and \ropt{2} introduces new communication edges to force an increase in distortion.
What is left to show is how the value of $\dist_{\h}(E_i)$ comes into play, and which edges the adversary commits to.

Note that only $n-1$ edges can be revealed in total (since the final request graph is a line), and that an offline algorithm incurs a cost of at most $n$ to lower the distortion of an edge to $1$.
Thus, in order to prove \rthm{lower} the adversary must -- on average -- be able to force a cost of $\Omega(n \log n)$ per edge revealed.
As will be apparent from our construction, the factor $\log n$ comes from the way the adversary reveals edges:
it first reveals $n/2$ edges, then $n/4$, $n/8$, etc.\ , resulting in $\log n$ batches.
After each batch, for \ON{} to remain optimal it must permute its layout at cost $\Omega(n^2)$, totalling a cost of $\Omega(n^2 \log n)$ for all batches combined.
\color{black}
To ensure that $R(\sigma_\ON)$ is a line graph, the partial request graph $E_i$ (i.e.\ the set of revealed edges) always comprises a set of disjoint \emph{sublists}.
Therefore, the adversary only reveals edges that concatenate two sublists in $E_i$.
Initially $E_i$ is empty and the corresponding sublists are all singleton sets of $u \in V$.

\begin{figure}[h]
\center
\includegraphics[width=.65\textwidth]{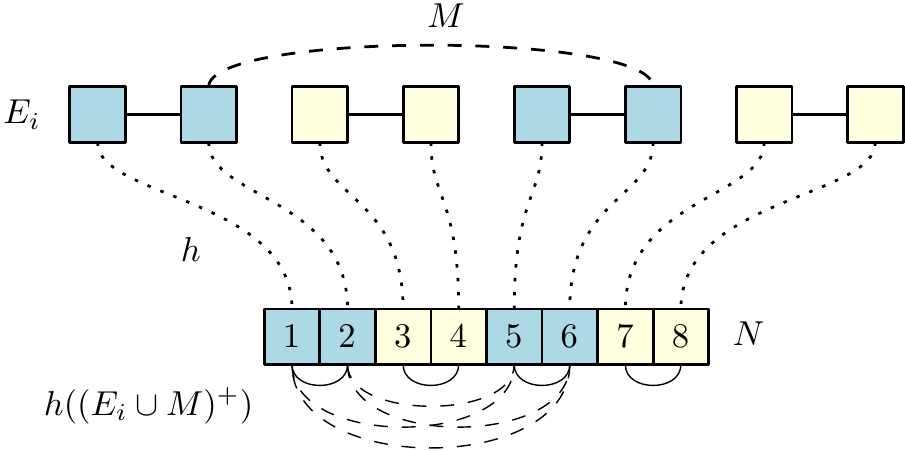}
\caption{A visualization of $\dist_{\h}(E_i \cup M)$: the line graph N, $E_i$ (solid) and $M$ (dashed) are sets of edges, configured on $N$ by $h$ (dotted). The sum of length of the configured edges $h((E_i \cup M)^+)$ is the distortion $\dist_{\h}(E_i \cup M)$.}
\label{fig:distortion}
\end{figure}

To help decide which edges to reveal, we use the distortion to associate a cost to batches of edges that the adversary can commit to.
Let $M \subseteq V \times V \setminus E_i$ be any set of edges such that the graph $(V, E_i \cup M)$ comprises a set of disjoint sublists.
For a configuration $\h$ of \ON, the set $M$ induces a distortion of $\dist_{\h}(E_i \cup M)$, as shown in \rfig{distortion}.
We show that for any embedding that \ON{} chooses, the adversary can find a set $M$ so that the distortion is large.
To formalize this, we prove the following.

\begin{lemma}
\label{lem:linematching}
Let $N$ be a line graph, and $E \subseteq V\times V$ a set of edges so that the graph $G=(V,E)$ induces $k$ disjoint sublists.
For every $\h \in \conf{V}{N}$, there exists a set $M \subseteq V \times V$ of at most $k/2$ edges such that $\dist_{\h}(h(E \cup M)) = \Omega(\frac{n^3}k)$
and $(V, E \cup M)$ comprises a set of disjoint lists.
\end{lemma}

\color{red}
\begin{proof}
Let $L_1,...,L_{k} \subseteq E$ be the sublists in $G$.
For all pairs $(i, j)$, let $(L_i, L_j)$ denote any edge so that $L_i \cup L_j \cup \{(L_i, L_j)\} =L_i\oplus L_j$ is connected.
For any involution\footnote{A function that is its own inverse, i.e.\ $f(f(i)) = i$.} $f$ on the sublists we have:
\begin{equation}
2 \dist_{\h}(E \cup \{(L_i, L_{f(i)}) \mid i \not = f(i)\}) \geq \sum_{i=1}^{k} d_\h(L_i\oplus L_j).
\label{eq:1}
\end{equation}
The factor $2$ is necessary because for $i$ such that $i \not = f(i)$, the term $d_\h(L_i\oplus L_{f(i)})$ appears twice to the sum.

Now partition $N$ into three sublists: a left part $X = \{1,\ldots,\ceil{n/3}\}$, a right part $Y = \{\floor{2n/3}, \ldots, n\}$, and the centre part $C = N \setminus (X \cup Y)$.
Let $\h_X(L_i)$ (resp. $\h_Y(L_i)$) denote the number of elements of $L_i$ that $\h$ maps onto $X$ (resp. $Y$).
Every two vertices $u,v$ so that $\h(u) \in X$ and $\h(u) \in Y$ are by construction at least $|C| = \Theta(n)$ apart on $N$,
and therefore we can lower bound $d_\h(L_i \oplus L_j)$ by:
\begin{equation}
d_\h(L_i \oplus L_j) \geq |C|\cdot\h_X(L_i)\h_Y(L_j)
\label{eq:2}
\end{equation}

For an involution $f$ drawn uniformly at random, \rthm{staircase} gives us a bound on the expected value of the following:
\begin{equation}
\mathbf{E} \left(\sum_{i=1}^{k}\h_X(L_i)\h_Y(L_{f(i)})\right) = \Omega\left(\frac{\ceil{n/3}^2}{k}\right)
\label{eq:3}
\end{equation}
Therefore, there exists an involution $f$ for which we have:
\begin{align*}
2\dist_{\h}(E \cup \{(L_i, L_{f(i)}) \mid i \not = f(i)\}) & \overset{(\ref{eq:1})}{\geq} \sum_{i=1}^{k} d_\h(L_i + L_{f(i)})\\
 &\overset{(\ref{eq:2})}{\geq} |C|\cdot\sum_{i=1}^{k} \h_X(L_i)\h_Y(L_{f(i)})\\
 &\overset{(\ref{eq:3})}{=} \Theta(n)\cdot\Omega(n^2 / k) =  \Omega\left(\frac{n^3}k\right)
\end{align*}

Since this holds for any choice of $(L_i, L_j)$, we can pick them so that $(V, E \cup \{(L_i, L_{f(i)}) \mid i \not = f(i)\})$
comprises a set of disjoint lists. \qed



\end{proof}

\color{black}

\textcolor{red}{
This lemma (and the proof) reveals how the adversary commits to a new batch of edges in \ropt{2} (essentially a random matching will do).}
Observe that the number of edges is at most half the number of sublists in $E_i$.
In the worst case we have to assume it is exactly half, and thus that the number of sublists is halved after every new batch of edges is selected.
Next we show the precondition for the adversary to opt for \ropt{1}, including a lower bound on the corresponding cost imposed on \ON.

\begin{lemma}
\label{lem:precondition}
Let $N$ be a line graph, $\h \in \conf{V}{N}$ a configuration, and $E \subseteq V \times V$ a set of edges so that the graph $G=(V,E)$ has $n/\ell$ disjoint sublists of size $\ell$.
If $\dist_{\h}(E) = \Omega(\ell n^2)$, then there exists an edge $(u,v) \in E$ such that $\dist_{\h}(u,v) = \Omega(n/\ell)$.
\end{lemma}

\begin{proof}
There are at most $n/\ell \cdot {\ell \choose 2} = \Oh(\ell n)$ distinct simple paths in $G$, meaning that the average distortion of these paths is $\frac{\Omega(\ell n^2)}{\Oh(\ell n)} = \Omega(n)$.
The highest distortion is at least the average, and every path in $G$ has length at most $\ell$.
On this path, there must exist an edge with distortion $\Omega(n/\ell)$, 
since if all edges have a distortion of $o(n/\ell)$, the total would be $o(n)$.\qed
\end{proof}

Combined, \rlem{linematching} and \rlem{precondition} imply that the adversary can either request an edge at cost $\Omega(n/\ell)$, or increase the distortion to $\Omega(\ell n^2)$ by revealing a new batch of edges.
The final ingredient is a lower bound on how much cost the adversary can impose on \ON{} in between these batches.

\begin{lemma}
\label{lem:halfdisplacement}
Let $N$ be a line graph, $E \subset V \times V$ a set of communication edges.
If $\h, \h' \in \conf{V}{N}$ are two embeddings that differ only in the order of two adjacent elements $u$ and $v$,
then $\dist_{\h}(E) \leq \dist_{\h'}(E) + 2\ell$, where $\ell$ is the size of the largest sublist in~$E$.
\end{lemma}

\begin{proof}
Consider all simple paths in $E$ that end in $u$.
At most $\ell$ paths ending in $u$ are reduced by 1, and similarly at most $\ell$ paths ending in $v$.
Therefore $\dist_{\h}(E)) - \dist_{\h'}(E) \leq 2\ell$.
\end{proof}

Combining the previous lemmata, we can prove the main technical result.

\begin{lemma}
\label{lem:linebound}
For every online algorithm A, there is a sequence $\sigma_\ON$ of length $\Oh(\varepsilon n^{1+\varepsilon}  \log n)$ such that $\Cost(\ON(\sigma_\ON)) = \Omega (\varepsilon n^2  \log n)$, for $0 < \varepsilon \leq 1$.
Furthermore, the resulting request graph $R(\sigma_\ON)$ is a line graph.
\end{lemma}

\begin{proof}
W.l.o.g.\ assume that $n = 2^p$ for some integer $p$.
This implies that the number of edges in every new batch is a power of $2$; consequently, the sublists in any set $E_i$ of revealed edges have size $2^k = \ell$ for some integer $k$.

Consider the situation right after a batch of edges is revealed, where all sublists have size $\ell$.
By \rlem{linematching} this implies that the distortion is $\Omega(\ell n^2)$.
Let $\sigma = \sigma_i,\sigma_{i+1},...,\sigma_{i + \ell n}$ be the requests obtained by repeatedly requesting the edge in $E_i$ with largest distortion.
There are two situations:
\begin{itemize}
    \item Throughout serving $\sigma$, the distortion is always at least $\Omega(\ell n^2)$.
    Then by \rlem{precondition} each $\sigma_j$, $i \leq j \leq i+ \ell n$ incurred a cost of $\Omega(n/\ell)$, at total cost $\Omega(n^2)$.
    \item By serving $\sigma$, \ON{} halves the distortion, thus reducing it by at least $\Omega(\ell n^2)$.
          Then, since by \rlem{halfdisplacement} every swap reduces the distortion by at most $2\ell$, \ON{} must have used at least $\Omega(n^2)$ swaps.

\end{itemize}

This argument holds for each batch of edges revealed.
The adversary stops when the sublists have size $2^{\varepsilon \log n}$, yielding a sequence $\sigma_\ON$ of length $\Oh(\varepsilon n^{1+\varepsilon} \log n)$ with a cost of $\Omega(\varepsilon n^2 \log n)$ for \ON.
By \rlem{precondition}, the adversary only requests edges that are introduced using the matching from \rlem{linematching}.
Any edge introduced by the latter Lemma concatenates two already existing sublists, hence $R(\sigma_\ON)$ is a line graph. \qed
\end{proof}

To wrap up the proof for \rthm{lower}, we conclude by showing that for any online algorithm \ON, the sequence $\sigma_\ON$ can be solved in $\Oh(n^2)$ by an optimal offline algorithm.

\begin{proof}[Proof of \rthm{lower}]
Let $\ON$ be any online algorithm solving \pwl{}.
Apply \rlem{linebound} with $\varepsilon = 1/2$, yielding $\Cost(\ON(\sigma_\ON)) = \Omega(n^2 \log n)$.
Since $\sigma_\ON$ is a line graph, an offline algorithm can embed this graph at (worst case optimal) cost $\Theta(n^2)$, and serve every request at optimal cost $\Oh(1)$.
This yields $\Cost(OFF(\sigma_\ON)) = \Theta(n^2)$, and thus
\[\rho = \frac{\Cost(\ON(\sigma))}{\Cost(\OFF(\sigma))} = \Omega(\log n)\]

In order to make this bound hold for arbitrary long sequences, we slightly modify the adversary.
After every $\Oh(n^2)$ requests it serves, it can reconfigure to a new list at cost $\Oh(n^2)$, and repeat the argument to force cost of $\Omega(n^2 \log n)$ to \ON{} for the subsequent $\Oh(n^2)$ requests.
\end{proof}

\noindent \textbf{Remark.} We can extend the model for \pwl{} to include cases where both the communication graph and the host graph $G$ are a $d$-dimensional grid, for constant $d$; we dub this problem \dga{}.
On a request $(u,v)$, the cursor is placed at $u$ and the request is served when it touches $v$.
The same operations are allowed: \textbf{moving} the cursor, or \textbf{swapping} with on of its $2^d$ neighbors (also moving the cursor).

We can extend our lower bound to \dga{}.
That is, we can construct a sequence $\sigma_\ON$ such that $R(\sigma_\ON)$ is a $d$-dimensional grid.
We show that an offline algorithm can perfectly embed this graph at cost at most $n^{1+1/d}$, whereas we can force the online algorithm to permute its layout $\log n$ times.

What follows here is a summary of the necessary changes to the argument presented in the previous section.
In particular, how to prove the following generalized version of \rlem{linebound}.

\begin{lemma}
\label{lem:gridbound}
For every online algorithm \ON{} for \dga{}, there is a sequence $\sigma_\ON$ of length $\Oh(\varepsilon n^{1+\varepsilon}  \log n)$ such that $\Cost(\ON(\sigma_\ON)) = \Omega (\varepsilon n^{1+1/d} \log n)$, for $0 < \varepsilon \leq 1$.
The resulting request graph $R(\sigma_A)$ is a $d$-dimensional grid graph.
\end{lemma}

First some semantic changes to the notation.
In all lemmata, we interpret $N$ to be a $d$-dimensional grid, where the number of vertices is $n$.
The distance $\dist_{\h}(u,v)$ for $u, v \in V$ can then be interpreted as the $\ell_1$ norm.

Second, two minor technical changes to the proofs.
The difference with \rlem{linebound} is that the diameter of a $d$-dimensional grid with $n$ nodes is at least $n^{1/d}$,
which is exactly the value $|C|$ in the proof of \rlem{linematching}.
The rest of the arguments almost directly generalize to yield \rlem{gridbound}.
The final detail to be careful about is that the \emph{length} of the sequence $\sigma_\ON$ does not exceed the \emph{cost} of the offline algorithm.
That is, we have to pick $\varepsilon$ such that $n^{1+\varepsilon}  \log n = \Oh(n^{1+1/d})$.
For constant dimension $d$, we can achieve this by picking $\varepsilon = \frac 1 {2d}$.

\section{An Upper Bound}\label{sec:upperbound}

This section presents a $\Oh(\log n)$-competitive online algorithm for \pwl{}.
Our main technical lemma shows that the total cost spent on learning the optimal embedding never exceeds $\Oh(n^2 \log n)$.
We propose a simple greedy algorithm that identifies a \emph{locally optimal} embedding, and always moves towards this embedding.
Let $N$ be a line graph, and $\h \in \conf{V}{N}$ a configuration.
An $\h$-optimal embedding of $E \subseteq V \times V$, denoted $\opt{h}{E}$, is an embedding that optimally embeds every connected component of $E$ while minimizing the quantity $\sum_{v \in V} |\h(v) -\h'(v)|$.
That is, it is the optimal embedding of $E$ that is `closest' to $h$.
With such a configuration we associate the cost:
\[\Phi_{\h}[E] = \sum_{v \in V} |h(v) - \opt{h}{E}(v)|\]
Let \GR{} be the algorithm (it GREedily ADjoins sublists), that upon seeing a new edge $\sigma_i$, \emph{immediately} moves to the embedding $\opt{\h}{E_i \cup \{\sigma_{i+1}\}}$.

\begin{figure}
\center
\includegraphics{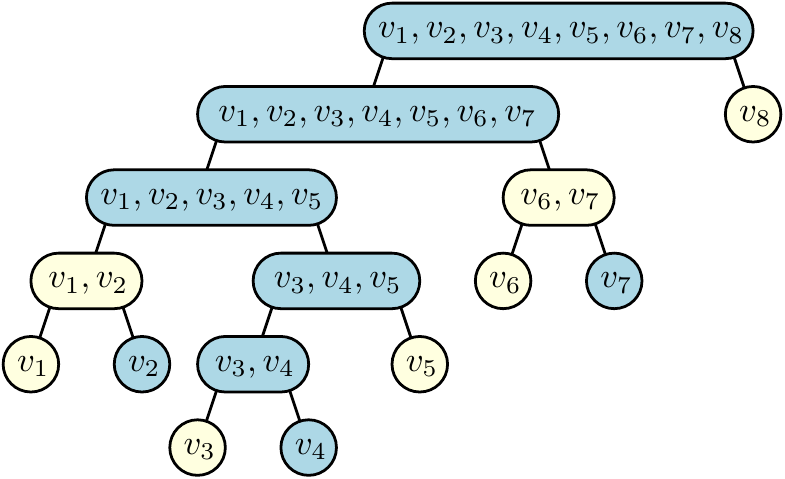}
\caption{The tree $(\mathcal{V}_\sigma, E_\sigma)$ for $\sigma =(v_1, v_2), (v_3, v_4), (v_5, v_3), (v_6, v_7), (v_1, v_5), (v_4, v_7), (v_8, v_2)$.
The smallest and largest subtrees have light and dark backgrounds respectively.}
\label{fig:mergetree}
\end{figure}

For each $E_i$, let $\mathcal{V}(E_i)$ be the connected components of $(V, E_i)$, so that $\mathcal{V}_\sigma = \cup_{1\leq i \leq m} \mathcal{V}(E_i)$ is the set of all sublists induced by $\sigma$.
This naturally defines a binary tree $T_\sigma = (\mathcal{V}_\sigma, E_\sigma)$:
for every first occurence $\sigma_i$ of $(u,w) \in E_m$ connecting two sublists $U, W$ in $R(E_i)$, there are two corresponding edges $(U, U \cup W)$ and $(W, U\cup W)$ in $E_\sigma$ (see \rfig{mergetree}).
For every $\sigma_i \in E_m$, \GR{} incurs some cost for reconfiguring, and the following lemma bounds this cost.

\begin{lemma}
\label{lem:reconfigure}
Let $E_i$ be as before and let $\sigma_{i} \in E_m$ be an edge connecting two sublists $U$ and $W$ of $E_{i-1}$.
It holds that
\[\Phi_h[E_i \cup \{\sigma_{i+1}\}]-\Phi_h[E_i] \leq n \cdot \min (|U|, |W|)\]
\end{lemma}

\begin{proof}
With $\Phi_h[E_i]$ moves, we can optimally embed $E_i$.
With an additional $n\cdot \min(|U|,|W|)$ moves, we can relocate the smaller of $|U|$ and $|W|$ to achieve an optimal embedding of $E_i \cup \{\sigma_{i+1}\}$.
Therefore $\Phi_h[E_i \cup \{\sigma_{i+1}\}] \leq \Phi_h[E_i] + n \min (|U|, |W|)$, and the claim follows.
\end{proof}

For a node $U \in \mathcal{V}_\sigma$, let $\mathrm{left}(U)$ and $\mathrm{right}(U)$ denote $U$'s left and right child respectively.
Further, let $w(U)$ denote the number of nodes in the subtree rooted at $U$.
Observe that for any binary tree with nodes $N$, it holds that
\[\sum_{v \in N} \min(w(\textrm{left}(v)), w(\textrm{right}(v))) \leq |N| \log |N|\]

\begin{theorem}
\label{thm:greed}
For any $\sigma$, with $|\sigma| = m$, such that $|E_m| = k$ and $R(\sigma)$ is a line graph,
\[\Cost(\GR(\sigma)) = \Oh(m + nk \log k)\]
\end{theorem}

\begin{proof}
The total cost of \GR{} is the sum of reconfiguring after every $\sigma_i \in E_m$ plus accessing every request at cost 1:
\begin{align*}
\Cost(\text{\GR}(\sigma)) - m &= \sum_{\sigma_i \in E_m} \Phi_h[E_i \cup \{\sigma_{i}\}]-\Phi_h[E_i]]\\
&\leq \sum_{U \in \mathcal{V}_\sigma} n \min(w(\textrm{left}(U)), w(\textrm{right}(U))) \\
&\leq n k \log k
\end{align*}
\end{proof}

As a corollary, it is not hard to show that \GR{} achieves optimal $\log n$ competitiveness for the worst case sequence constructed in \rsec{lowerbound}.
Additionally, in Appendix \ref{sec:distributed} we show a distributed implementation of this algorithm using message passing.

\section{Related Work}\label{sec:relwork}

As discussed above, the motivation for our work stems
from the increasing flexibilities available in networked systems,
supporting resource migrations and reconfigurable topologies. 
In the following, we will review works related to the technical and algorithmic
contributions in this paper. 

One important area of related work arises in the context of
the dynamic list update problem. Since the groundbreaking work 
by Sleator
and Tarjan on amortized
analysis and self-adjusting datastructures~\cite{sleator1985amortized},
researchers have also explored many interesting variants
of self-adjusting datastructures, also using randomized 
algorithms~\cite{reingold1994randomized} or lookaheads~\cite{albers1998competitive,albers1995combined},
or offline algorithms~\cite{ambuhl2000offline,reingold1996off}.
The deterministic Move-To-Front (MTF) algorithm is known to optimally solve
the standard formulation of the list update problem: it is 2-competitive~\cite{sleator1985amortized}, 
which matches the lower bound~\cite{albers1998self}. 
To the best of our knowledge, the competitive ratio in the randomized
setting (against an oblivious adversary) is still an open problem: 
the best upper bound so far is
1.6~\cite{albers1995combined}, and the best 
lower bound 1.5~\cite{teia1993lower}.  
The randomized algorithm~\cite{albers1995combined} makes an initial random 
choice between two known algorithms that have different worst-case 
request sequences, relying on the BIT~\cite{reingold1994randomized} 
and TIMESTAMP~\cite{albers1998improved} algorithms.

We also note that the self-adjusting linear network design problem
can be considered a special case of general online problems
such as the online Metrical Task System (MTS) problems.
However, given the exponential number of possible configurations,
the competitive ratio of generic MTS algorithms will be high
if applied to our more specific problems (at least according to
the existing bounds). Furthermore, we note that in case of line
request graphs, the problem can also be seen as a learning
problem and hence related to bandits theory~\cite{bubeck2012regret}. 

In terms of reconfigurable networks, 
there exist several static~\cite{rdan,infocom19dan} and dynamic~\cite{ton15splay,infocom19splay}
algorithms for constant-degree networks, as well as
hybrid variants~\cite{ifip19dan} which combine static and reconfigurable
links. However, these solutions do not apply to the line
and do not provide performance guarantees over time;
the latter also applies to recent work on node migration models
on the grid~\cite{avin2013locally}.

The paper closest to ours is by Olver et al.~\cite{olver2017itinerant}
who introduced the Itinerant List Update (ILU) problem:
a relaxation of the classic dynamic list update problem in 
which the pointer no longer has to return to a home location after
each request. The authors show an $\Omega(\log{n})$
lower bound on the randomized competitive ratio and
also present an offline polynomial-time algorithm 
and prove that it achieves an approximation ratio of
$O(\log^2 n)$. In contrast, we in our paper focus on
online algorithms and request graphs forming a line
(or grid). In fact, we show that the lower bound $\Omega(\log{n})$
even holds in this case, at least for deterministic algorithms.
We also present an online algorithm which matches this bound
in our model. 

\section{Conclusion}\label{sec:conclusion}

We presented a first and asymptotically tight, i.e., $\Theta(\log{n})$-competitive 
online algorithm for self-adjusting reconfigurable line networks with linear demand. 
Both our lower and upper bounds are non-trivial, and 
we believe that our work opens several interesting
directions for future research. In particular, it
would be very interesting to shed light on the competitive ratio
achievable in more general network topologies,
and to study randomized algorithms.

{
  \bibliographystyle{splncs04} 
\bibliography{literature}
}


\appendix

\section{Geometric Proof}

\begin{theorem}
\label{thm:staircase}

Let $x_1,\dots,x_k$ and $y_1,\dots,y_k$ be sequences of $k$ nonnegative numbers, and let $x$ (resp. $y$) denote $\sum_{i=1}^k x_i$.
Let the \emph{weight} of an \emph{involution}\footnote{A function $f$ such that $f(f(x)) = x$ for all $x$.} over the indices $1,\dots,k$ be defined as:
\[w(f) = \sum_{i = 1}^k x_iy_{f(i)}\]

The average weight over all involutions is $\Omega(\frac{xy}{k})$.
\end{theorem}

\begin{figure}[ht]
\center
\includegraphics{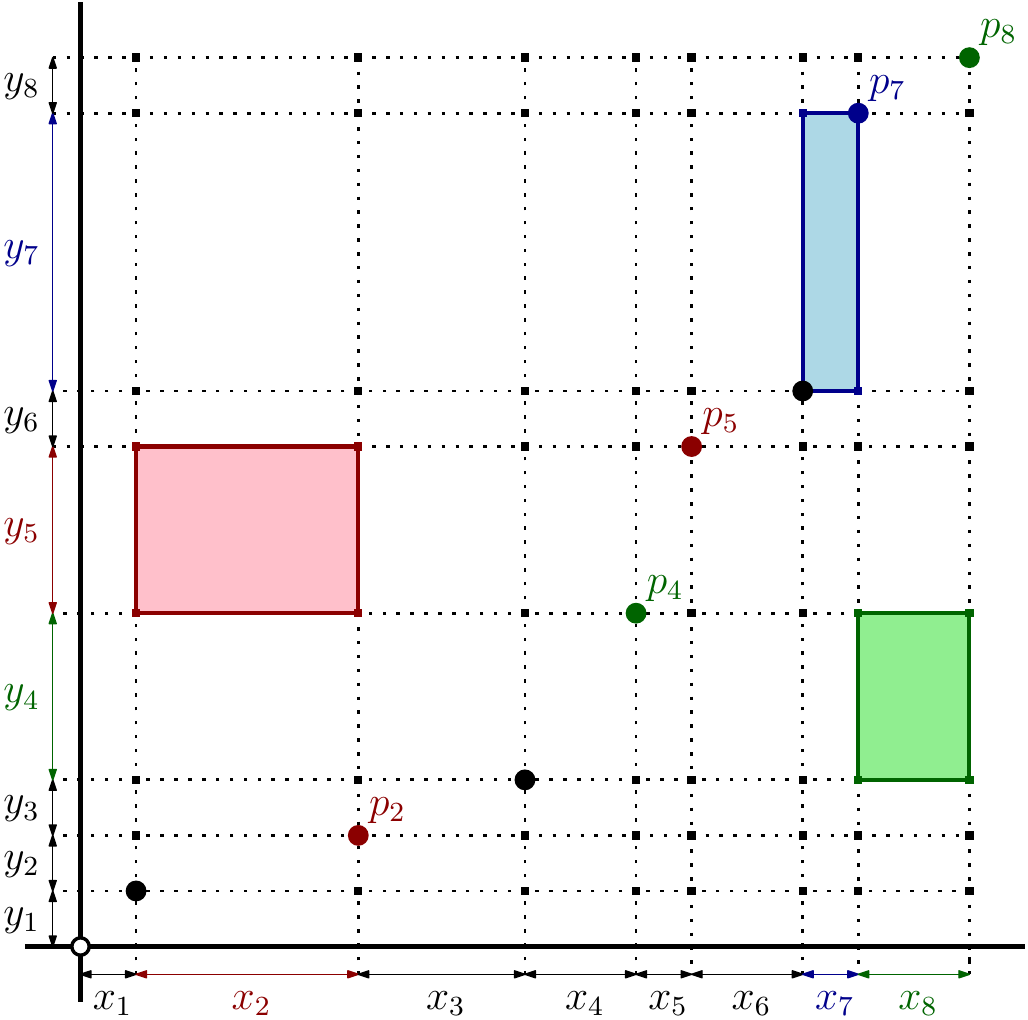}
\caption{A staircase of 8 points based on the sequences $x_1,\dots,x_8$ and $y_1,\dots,y_8$.
The values $w(2, 5)$, $w(8, 4)$, and $w(7, 7)$ are visualised as the area of rectangles highlighted in red, green, and blue respectively.}
\label{fig:staircase}
\end{figure}

\begin{proof}
Let $\mathcal{I}_k$ denote the set of all involutions on a set of $k$ elements, where $|\mathcal{I}_k| = T(k)$
is given by the recurrence $T(k) = T(k-1) + (k - 1)T(k-2)$ with $T(0) = T(1) = 1$.
For every pair of distinct indices $i, j$, there are $T(k-2)$ involutions $f \in \mathcal{I}_k$ such that $f(i) = j$
(namely for all involutions on the remaining $k-2$ indices).
Similarly for every index $i$, there are $T(k-1)$ involutions such that $f(i) = i$.
Thus, for every ordered pair of (not necessarily distinct) indices $i, j$ there are \emph{at least} $T(k-2)$ involutions with $f(i) = j$.

For convenience we define a \emph{staircase} of points $p_j = (\sum_{i=1}^j x_i, \sum_{i=1}^j y_i)$.
Observe that we can subdivide the rectangle defined by $p_k$ and the origin into $k^2$ axis-aligned rectangles,
so that the area of every such rectangle corresponds to the weight of one ordered pair of indices (see \rfig{staircase}).
Since every ordered pair of indices appears in at least $T(k-2)$ involutions, their weight (and thus the corresponding rectangle),
contributes at least $T(k-2)$ times in the sum of weights over all involutions.
This means that the area $xy$ of the complete rectangle contributes $T(k-2)$ times to that sum:
\[\frac{\sum_{f \in \mathcal{I}_k} w(f)}{|\mathcal{I}_k|} \geq \frac{T(k-2)\cdot \sum_{i=1}^k\sum_{j=1}^k w(i, j)}{T(k)} = \frac{T(k-2)}{T(k)} \cdot x y\]

To lower bound $\frac{T(k-2)}{T(k)}$, we first define $R(n) = \frac{T(n)}{T(n-1)}$
and observe that this definition is equivalent to the one in \rlem{rootbound}:
\begin{align*}
R(n) &= \frac{T(n)}{T(n-1)}\\
     &= \frac{T(n-1) + (n-1)T(n-2)}{T(n-1)}\\
     &= 1 + (n-1) \frac{T(n-2)}{T(n-1)}\\
     &= 1 + \frac{n-1}{R(n-1)}
\end{align*}
Since $\frac{T(n)}{T(n-2)} = R(n)R(n-1)$, we can use \rlem{rootbound} to lower bound $\frac{T(k-2)}{T(k)}$ by:
\[\frac{T(k-2)}{T(k)} = \frac{1}{R(k)R(k-2)} \geq \frac{1}{(1 + \sqrt{k+1})(1+\sqrt{k-1})} = \Theta\left(\frac1k\right)\]
thus yielding an average weight of $\Theta(\frac{xy}{k})$ over all involutions.
\end{proof}

\begin{lemma}
\label{lem:rootbound}
Let $R(n) = 1 + \frac{n - 1}{R(n-1)}$ with $R(1) = 1$; for all $n \geq 1$ it holds that:
\[\sqrt{n} \leq R(n) < 1 + \sqrt{n + 1}\]
\end{lemma}

\begin{proof}
The proof is by induction on $n$, with base case $\sqrt{1} \leq R(1) < 1 + \sqrt{1+1}$.
From $R(n) < 1 + \sqrt{n + 1}$ we conclude:
\[\sqrt{n + 1} = 1 + \frac{n}{1 +\sqrt{n + 1}} < 1 + \frac{(n + 1) - 1}{R(n)} = R(n + 1) \]
And from $\sqrt{n} \leq R(n)$ we conclude:
\[R(n + 1) = 1 + \frac{(n + 1) - 1}{R(n)} \leq 1 + \frac{n}{\sqrt{n}} = 1 + \sqrt{n} < 1 + \sqrt{(n+1) + 1}\]
\end{proof}

\section{Distributed Implementation of \GR{}}
\label{sec:distributed}
To make \GR{} distributed we have several problems to overcome: i) routing, ii) knowing to which (temporary) sublist every node belongs together with the size of the sublist, and iii) how to perform the reconfiguration and merging of two sublists. We address these issues one by one.

\textbf{Routing}: The basic problem with routing is that the source nodes do not know the location of the destination, since initially there is no \emph{sense of direction}. To overcome this problem each source initiates an exponential search on \emph{both} sides of the line network when it first needs to communicate with a destination. This will guarantee that the cost of the \emph{first} route request will be $O(i)$ for a destination that is $i$ hops away on the line network. Note that this is proportional  the cost of any algorithm. According to \GR{} the cost of all future requests will be 1 since after the first communication request the source and destination are reconfigured to be neighbors.

\textbf{Sublist}: During the execution each node maintains the following information: A bit that indicates if it is at the \emph{end} of a sublist (a node is at the end of a sublist if it has less than two neighbors from that sublist). If it is at the end of the list then the node maintains the size of the list (up to $\log n$ bits).

\textbf{Reconfiguration}: Basically \GR{}  merges two sublists by swapping the shorter list toward the longer sublist. Note that this happens only on the \emph{first} routing request from a source to destination. This can be done in a distributed manner in the following way. On the first routing request, the source (which must be an \emph{end} node) attaches the size of its sublist to the message. The destination (which also must be an \emph{end} node), upon receiving the request,  answers to the source with the size of its own sublist (initially set to one).  It is then clear to both the source and destination which sublist needs to move toward which sublist and what will be the size of the merged sublist. Then, both source and destination send messages within their sublist informing the other \emph{ends} of the sublist of the size of the merged list.  Now, w.l.o.g assume the destination needs to move toward the source. The destination then starts performing swaps (with its neighbor that is not on its current list) toward the source. This process ends when both the destination is a neighbor of the source and the source is a neighbor of its previous neighbor on its list. Before starting the swaps the destination informs its neighbor (which in turn informs its neighbor and so on) to \emph{follow up} after it with similar swaps. It can be observed that after this process the two list will be merged into a larger list and both \emph{ends} will know the sizes of the new sublist. The cost of the reconfiguration is $O(n \min (|U|, |W|))$ where $U$ and $W$ are the two sublists involved in the merging.

\end{document}